\documentclass[%
reprint,
amsmath,amssymb,
aps,
]{revtex4-1}

\usepackage{graphicx}% Include figure files
\usepackage{dcolumn}% Align table columns on decimal point
\usepackage{bm}% bold math
\usepackage{amsmath}
\usepackage{enumerate}
\usepackage{setspace}
\usepackage{dsfont}
\usepackage{subfigure}
\usepackage{multirow}

\usepackage{threeparttable}
\usepackage[pdfstartview=FitH,
CJKbookmarks=true,
bookmarksnumbered=true,
bookmarksopen=true,
colorlinks, 
pdfborder=001,
linkcolor=blue,
anchorcolor=blue,
citecolor=blue
]{hyperref}

\begin{document}

	\title{An inhibited laser}
	
	\author{Tiantian Shi$^{1}$, Duo Pan$^{1}$  \& Jingbiao Chen$^{1,*}$\footnote[*]{E-mail: jbchen@pku.edu.cn}}
	
	\address{$^1$State Key Laboratory of Advanced Optical Communication Systems and Networks, Institute of Quantum Electronics, School of Electronics, Peking University, Beijing 100871, China}

	\begin{abstract}
	
	Traditional lasers function using resonant cavities, in which the round-trip optical path is exactly equal to an integer multiple of the intracavity wavelengths to constructively enhance the spontaneous emission rate. By taking advantage of the enhancement from the resonant cavity, the narrowest sub-10-mHz-linewidth laser and a 10$^{-16}$-fractional-frequency-stability superradiant active optical clock (AOC) have been achieved. However, a laser with atomic spontaneous radiation being destructively inhibited in an anti-resonant cavity, where the atomic resonance is exactly between two adjacent cavity resonances, has not been reported. Herein, we experimentally demonstrate inhibited stimulated emission and termed it an inhibited laser. Compared with traditional superradiant AOCs, which exhibit superiority in terms of the high suppression of cavity noise, the suppression of the cavity-pulling effect of an inhibited laser can be further improved by a factor of ${\left( {{{{\text{2}}\mathcal{F}} \mathord{\left/{\vphantom {{{\text{2}}\mathcal{F}} \pi }} \right.\kern-\nulldelimiterspace} \pi }} \right)^2}$, i.e., 2.07 in this work, which was improved from 26 to 53 times. This study will guide further development of AOCs with better stability, and thus, it is significant for quantum metrology and may lead to new research in the laser physics and cavity quantum electrodynamics fields. 
		
	\end{abstract}
	
	\date{\today}
	\maketitle

	\section*{Introduction}
	\label{Introduction}
	
	The significantly enhanced spontaneous decay rate of the spin in a resonant circuit, known as the Purcell effect \cite{Purcell1995Spontaneous}, was first reported by Purcell in 1946. It was practically observed in the 1980s using atoms in resonant cavities both in the microwave \cite{Goy1983Observation} and optical \cite{Heinzen1987Enhanced,Martini1987Anomalous} domains. The 
	enhanced spontaneous radiation has important application potential in cavity quantum electrodynamics (QED) \cite{Mabuchi2002Cavity,Wallraff2004Strong,YoshieVacuum2004}, including for one-atom lasers \cite{McKeeverExperimental2003}, ion-trap lasers \cite{MatthiasExperimental2004}, and quantum logic gates \cite{MonroeDemonstration1995} in quantum computers.

	Essentially, the resonant cavity, whose cavity-mode frequency resonates with the peak of the emission line for atomic transition, enhances the strength of vacuum fluctuations, which promotes the atomic spontaneous radiation. Conversely, the spontaneous decay rate is suppressed when the cavity is off resonance, which was first proposed by Kleppner in 1981 and demonstrated through inhibited blackbody absorption \cite{VaidyanathanInhibited1981} and inhibited spontaneous emission \cite{Kleppner1981Inhibited}. After, inhibited spontaneous emission was experimentally demonstrated in microwave and optical cavities in 1985 \cite{Hulet1985Inhibited} and 1987 \cite{Heinzen1987Enhanced}, respectively. Heinzen \cite{Heinzen1987Enhanced} pointed out that in an anti-resonant cavity, where the atomic frequency was exactly at the center of two adjacent cavity resonances, the inhibition of the atomic spontaneous decay rate was the greatest. More strikingly, through coupling with an anti-resonant cavity, the atomic radiative level shift vanished and the spectral linewidth narrowed \cite{Heinzen1987Vacuum}, which is potentially useful for precision measurements. Despite the experimental success of inhibited spontaneous emission, it remains to be explored whether laser oscillations can be realized when the spontaneous emission rate is suppressed to the greatest extent in an anti-resonant cavity. The characteristics of an inhibited laser are alternative to explore in depth for rich development in the fields of laser physics.

	Nevertheless, the demonstration of inhibited spontaneous emissions has provided credible evidence for the observation of inhibited stimulated emissions. The spontaneous emission can be viewed as a stimulated emission originating from the vacuum fluctuations, and the spontaneous emission below the threshold determines the spectrum of the laser above the threshold \cite{Loudon2000The}. It has significant potential to achieve inhibited lasing, with the aid of a three- or four-level structure to increase the pumping efficiency and the multi-atom system to reach the strong-coupling regime \cite{Fox2007Quantum}. 
	
	Different from the traditional types of lasers working in the resonant-cavity region, in this work, we propose a laser operating in the antiresonant-cavity regime, where the atomic gain line is located exactly at the center of two adjacent cavity resonances, which is termed an inhibited laser. The characteristics of the inhibited laser, such as the intracavity photon lifetime and primary laser power behavior, laser linewidth, and cavity-pulling effect, under the conditions of an anti-resonant cavity, were proven experimentally and theoretically. In particular, we proved that the frequency shift of the laser oscillation vanished in an inhibited laser with reduced sensitivity to the cavity-length fluctuations. Compared with resonant active-optical-clock (AOC) lasing \cite{Chen2009Active,Bohnet2012A,PhysRevA.87.013821,PhysRevLett.123.103601,Stefan2020PhysRevA,PRL2020Holland}, the influence of the thermal cavity-length noise on the frequency of an inhibited laser is further suppressed by a factor of ${\left( {{{{\text{2}}\mathcal{F}} \mathord{\left/{\vphantom {{{\text{2}}\mathcal{F}} \pi }} \right.\kern-\nulldelimiterspace} \pi }} \right)^2}$.

	\section*{Results}
	\label{Results}
	
\begin{figure}
		\centering
		\includegraphics[width=1\linewidth]{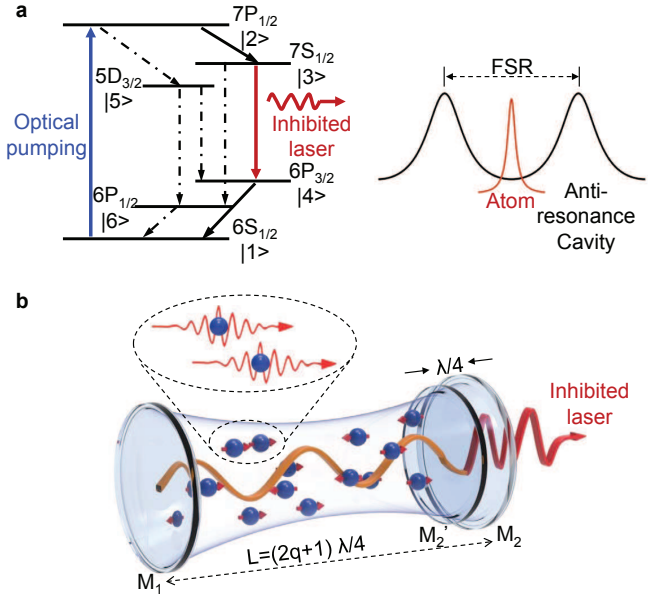}
		\caption{{\bf Working principle of the inhibited laser.} {\bf a,} Level scheme for $^{133}$Cs atom showing the 459 nm transition used for pumping and the 1470 nm transition as the inhibited lasing. Cs atoms are pumped by a 459 nm laser, which drives the $\left| 1 \right\rangle $ ($6{{\rm S}_{1/2}}$) to $\left| 2 \right\rangle $ ($7{{\rm P}_{1/2}}$) transition. Atoms are then transferred to the $\left| 3 \right\rangle $ ($7{{\rm S}_{1/2}}$) state through spontaneous radiation. Utilizing weak feedback of the cavity, population inversion is built up between the $\left| 3 \right\rangle $ and $\left| 4 \right\rangle $ ($6{{\rm P}_{3/2}}$) levels, and finally, 1470 nm lasing is realized. The inhibited laser is realized in an anti-resonant cavity, where the atomic gain line is exactly at the center of two adjacent cavity resonances. {\bf b,} Sketch of the inhibited laser. Cs atoms (blue spheres) along the direction of the cavity mode are pumped by the 459 nm laser. Two cavity mirrors are coated with reflectivities of ${{R}_1} = {{R}_2} = 34.5\%$ at 1470 nm. To change the cavity-mode frequency, the location of the output mirror is tunable. Generally, the round-trip optical path is exactly equal to an integer multiple of the intracavity wavelengths, and the output mirror is located at M$_{2}^{'}$. However, an inhibited laser (in red) is achieved when the round-trip optical path is equal to an odd multiple of the half wavelength $2L = (2q+1) \lambda /2$ with the output mirror being located at M$_2$.}
		\label{Fig1}
	\end{figure}

	{\bf Experimental scheme.} Here, we report an experimental demonstration of an inhibited laser. The general setup is depicted schematically in Fig. \ref{Fig1}, sharing similarities with the proposed superradiant AOC based on thermal atoms \cite{Shi2019OE}. $N \approx 1.8 \times {10^{11}}$ pure cesium (Cs) atoms are confined to the TEM$_{00}$ mode of a low-finesse optical cavity ($\mathcal{F}=3.07$), whose dissipation rate is $\kappa_0  = 2{\rm{\pi }}  \times 257$ MHz. Pumped by a 459 nm laser (6S$_{1/2}$-7P$_{1/2}$), the atoms achieve stimulated emissions at a wavelength of 1470 nm (7S$_{1/2}$-6P$_{3/2}$). The relaxation rate of the atomic dipole $\Gamma  = 2{\rm{\pi }}  \times 10.04$ MHz is much smaller than $\kappa_0$. Therefore, the laser works in a bad-cavity regime \cite{Chen2009Active,Bohnet2012A}. Unlike traditional resonant lasers, the inhibited laser is realized with a round-trip optical path equal to odd multiples of the half wavelength $2L = \left( {2q + 1} \right){\lambda  \mathord{\left/
			{\vphantom {\lambda  2}} \right.\kern-\nulldelimiterspace} 2}$. Information about the experimental details, as well as a discussion about the working regime of the laser, is provided in the Methods Sections I and II.

	{\bf Enhanced and inhibited factors.} Suppose that the atom emitting the first photon by spontaneous radiation is located at the center of cavity, and the reflectivities of cavity mirrors are ${R}_1 = {R}_2 = {R}$, the ratio of the power of spontaneous radiation emitted into cavity, ${P_{\rm{c}}}$, to the power into free space, ${P_{\rm{free}}}$, is given by \cite{Heinzen1987Vacuum}
	\begin{equation}\label{Eq1}
		\frac{{{P_{\rm{c}}}}}{{{P_{{\rm{free}}}}}} = \frac{{1 - {R^2}}}{{1 + {R^2} - 2R\cos \left( {{{\omega 2L} \mathord{\left/{\vphantom {{\omega 2L} {\rm{c}}}} \right.	\kern-\nulldelimiterspace} {\rm{c}}}} \right)}},
	\end{equation} 
	where $\omega$ is the angular frequency of the radiation, and c the speed of light. $\Delta \phi  = {{\omega 2L} \mathord{\left/
			{\vphantom {{\omega 2L} {\rm{c}}}} \right.
			\kern-\nulldelimiterspace} {\rm{c}}}$ denotes the phase shift of the intracavity reflected field, and it also indicates the detuning of the cavity frequency ${\omega _{\rm{c}}}$  from the atomic resonance $\omega_0$. A phase shift of ${\rm{2\pi }}$ between two consecutive round trips of the radiation inside the cavity corresponds to the cavity-frequency detuning ${\omega _{\rm{c}}}-\omega_0$ of one free spectral range (FSR). According to Eq. (\ref{Eq1}), the spontaneous decay rate from the atomic excited state is enhanced and inhibited by a factor of $\frac{{1 + R}}{{1 - R}}$ compared with that in free space when the cavity is resonant ($\Delta \phi  = {\rm{2}}\pi q $) and anti-resonant ($\Delta \phi  = \left( {2q + 1} \right){\rm{\pi }}$), respectively. Accordingly, the suppression of the spontaneous emission rate induced by the anti-resonant cavity is weak in the low-reflectivity case, which is conducive to realize inhibited lasing.

	\begin{figure*}[t]
		\begin{center}
			\includegraphics[width=1\linewidth]{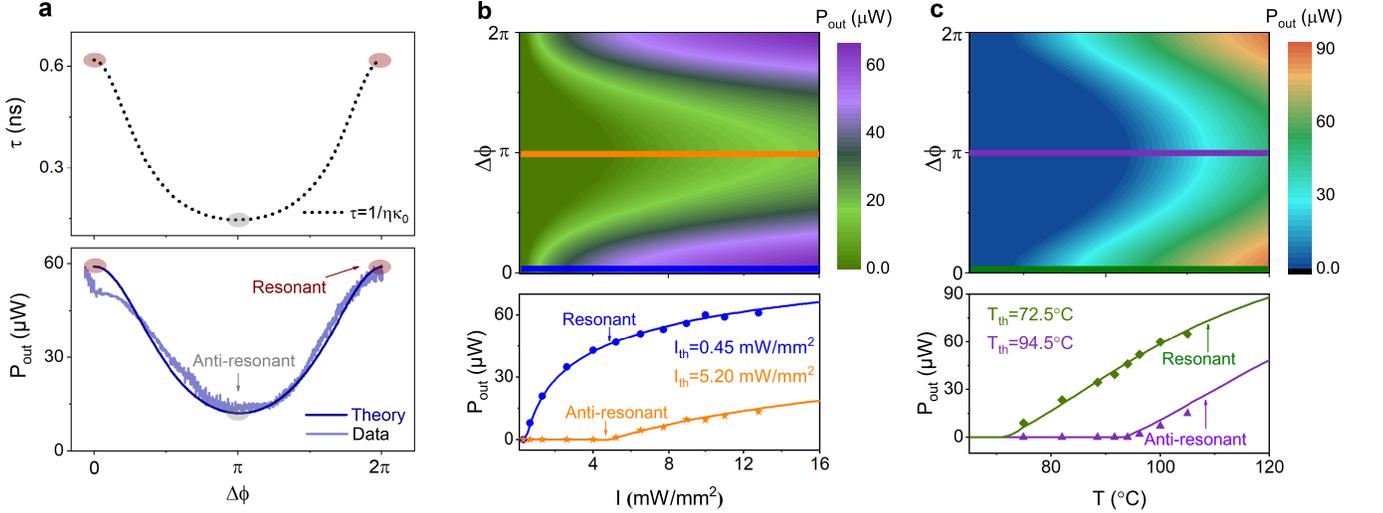}
			%vspace{-2mm}
		\end{center}
		\vspace{-5mm}
		\caption{{\bf Intracavity photon lifetime and primary laser power behavior.} {\bf a,} Intracavity photon lifetime $\tau $ (black dots) is inversely proportional to the loss coefficient $\eta $ described by Eq. (\ref{Eq3}). The experimental (light-blue line) and simulated (dark-blue line) results of the output laser power $P_{\rm{out}}$ match well under the condition of $g = 1.99 \times {10^5}\,{\rm{s}^{ - 1}}$, $\Omega  = 4.30 \times {10^7}\,{\rm{s}^{ - 1}}$, and $N_{\rm{eff}} = 5.71 \times {10^{9}}$. {\bf b,} Output laser power $P_{\rm{out}}$ as a function of the phase shift $\Delta \phi $ at different pumping light intensities $I$ under vapor-cell temperature $T = 100^{\circ}$C. The blue and orange lines represent $P_{\rm{out}}$ vs. the pumping light intensity when the cavity is on-resonant and anti-resonant, respectively. The blue circles and the orange stars are the corresponding experimental results. The pumping light intensities reaching the laser threshold were around 0.45 and 5.20 mW/mm$^2$ when the cavity was resonant and anti-resonant, respectively. {\bf c,} $P_{\rm{out}}$ as a function of $\Delta \phi $ at different vapor-cell temperatures $T$ under $I=10$ mW/mm$^{2}$. Theoretical and experimental $P_{\rm{out}}$ vs. $T$ when the cavity is resonant (green) and anti-resonant (purple) are separately depicted by the lines and dots, respectively. The corresponding temperatures for the laser threshold were 72.5$^{\circ}$C and 94.5$^{\circ}$C, respectively. }
		\label{Fig2}
	\end{figure*}

	{\bf Power characteristic of inhibited laser.} The detuning, $\Delta  = {\omega} - {\omega _0}$, of the radiation frequency $\omega$ from the atomic transition frequency $\omega_0$, can also be given by $\Delta  = P\left( {{\omega _{\rm{c}}} - {\omega _{\rm{0}}}} \right)$, where $P \equiv {{d\omega } \mathord{\left/
			{\vphantom {{d\omega } {d{\omega _{\rm{c}}}}}} \right.
			\kern-\nulldelimiterspace} {d{\omega _{\rm{c}}}}}$ represents the cavity-pulling coefficient \cite{Chen2009Active,Bohnet2012A}. In the bad-cavity limit, $P \approx {\Gamma  \mathord{\left/
			{\vphantom {\Gamma  {{\kappa _0}}}} \right.
			\kern-\nulldelimiterspace} {{\kappa _0}}} \ll 1$ if the cavity is near resonant, and thus, ${\Delta ^2} \ll 4{g^2}\left( {n + 1} \right)$ in this work. $n$ is the intracavity photon number, and the atom-cavity coupling constant \cite{PRA1997Kyungwon} $g = {\textstyle{\mu  \over \hbar }}\sqrt {\frac{{\hbar {\omega _0}}}{{2{\varepsilon _0}{V_{\rm{c}}}}}}  = 1.99 \times {10^5}{\mkern 1mu}$ s$^{-1}$, where $\mu$ is the electric dipole moment, $\varepsilon _0$ is the vacuum permittivity, and $V_{\rm{c}}$ is the equivalent mode volume. Consequently, the detuning $\Delta$ in the laser rate equation \cite{Christopher2002Multiple} is negligible (the exact 
	calculations are given in the Methods Section III). Here, we modify the loss term in the classical laser rate equation to obtain a universal expression, which can be used to describe any cavity-frequency detuning condition, as follows:
	\begin{align}\label{Eq2}
		\frac{{dn}}{{dt}} &= {N_{\rm{eff}}}\frac{{{\rho _{33}} - {\rho _{44}}}}{{{\tau_{{\rm{cyc}}}}}}{\sin ^2}\left( {\sqrt {n + 1} g{t_{{\rm{int}}}}} \right) - \frac{n}{\tau }
		\notag\\
		& = {N_{\rm{eff}}}\frac{{{\rho _{33}} - {\rho _{44}}}}{{{\tau_{{\rm{cyc}}}}}}{\sin ^2}\left( {\sqrt {n + 1} g{t_{{\rm{int}}}}} \right) - \eta \kappa_0 n.
	\end{align}
	On the right-hand side, the first term represents the gain, and the second term is the loss. For the gain term, the effective number of atoms that can be pumped to the 7P$_{1/2}$ state is $N_{\rm{eff}}=5.71 \times {10^{9}}$ with a pumping light intensity of $I=10$ mW/mm$^2$ and vapor-cell temperature of $T=100^{\circ}\rm{C}$. ${\rho _{ii}}$ denotes the population probability at level $\left| i \right\rangle $ in Fig. \ref{Fig1}a. ${\tau _{\rm{cyc}}}=28.0$ $\mu$s is the cycle time for Cs atoms through a transition of $6{{\rm S}_{1/2}} \to 7{{\rm P}_{1/2}} \to 7{{\rm S}_{1/2}} \to 6{{\rm P}_{3/2}}$. ${t_{{\rm{int}}}} \approx 19.8$ ns is the interaction time between the atoms and the cavity mode, and as a result, $g{t_{{\mathop{\rm int}} }} \ll {\rm{\pi }}$. The loss term is inversely proportional to the intracavity photon lifetime $\tau $. Typically, for the laser output from a resonant cavity, $\tau  = {1 \mathord{\left/
			{\vphantom {1 {{\kappa _0}}}} \right.
			\kern-\nulldelimiterspace} {{\kappa _0}}}$. However, if the cavity and the atomic-transition frequencies are not identical, $\tau  < {1 \mathord{\left/
			{\vphantom {1 {{\kappa _0}}}} \right.
			\kern-\nulldelimiterspace} {{\kappa _0}}}$. $\tau $ is exactly expressed as $\tau  = \frac{1}{{\eta \kappa_0 }}$. The loss coefficient $\eta$, reflecting the destructive interference of the intracavity radiated fields, is defined as the ratio of the maximum power emitted into the cavity at the resonant condition to the power at any cavity-frequency detuning,
	\begin{equation}\label{Eq3}
		\eta  = \frac{{{P_{{\rm{cmax}}}}}}{{{P_{\rm{c}}}}} = 1 + {\left( {\frac{{2\mathcal{F}}}{{\rm{\pi }}}} \right)^2}{\sin ^2}\left( {{{\omega L} \mathord{\left/
					{\vphantom {{\omega L} {\rm{c}}}} \right.
					\kern-\nulldelimiterspace} {\rm{c}}}} \right).
	\end{equation}
	Here, the approximation of $\mathcal{F} = \frac{{{\rm{\pi }}\sqrt R }}{{1 - R}}$ is used. As for the resonant cavity, the loss coefficient exhibits a minimum of ${\eta _{\min }} = 1$, and Eq. (\ref{Eq2}) is reduced to the traditional expression of the laser rate equation \cite{Christopher2002Multiple}. Instead, $\eta$ reaches the maximum ${\eta _{\max }} = 1 + {\left( {\frac{{2\mathcal{F}}}{{\rm{\pi }}}} \right)^2}$, which corresponds to the inhibited laser output from an anti-resonant cavity. The black dots in Fig. \ref{Fig2}a show the variation of $\tau$ with the change of $\Delta \phi$. 
	
	Using Eqs. (\ref{Eq2}) and (\ref{Eq3}), we obtain the steady-state solution of the intracavity photon number $n$ as a function of $\Delta \phi $ and further obtain the output laser power $P_{\rm{out}}$.  $P_{\rm{out}}$ as a function of of $\Delta \phi $ is represented by a dark-blue line in Fig. \ref{Fig2}a, and the experimental result is represented by the light-blue line. The deviation between the experimental and theoretical values was caused by ambient vibrations. This also confirmed that if the cavity was off-resonant, and the intracavity radiated fields interfered destructively, resulting in a decreased laser power. We measured $P_{\rm{out}}$ as a function of $\Delta \phi $ at different pumping light intensities $I$ and different vapor-cell temperatures $T$, as shown in Figs \ref{Fig2}b and \ref{Fig2}c, respectively. The power of the inhibited laser could be further improved with higher pumping light intensities and vapor-cell temperatures.

	\begin{table*}[htbp]
		\centering
		\caption{Cavity-pulling coefficient in optical domain}
		\label{table1}
		
		\begin{threeparttable}
			
			{\setlength{\tabcolsep}{18pt}
				\begin{tabular}{cccc }
					\hline
					\hline
					\multicolumn{4}{c}{Stimulated emission} \\ \hline
					\multirow{3}{*}{Resonant}      & Theoretical & Experimental & Simulated       \\ 
					\cline{2-4} 
					& $\frac{{{\Gamma}}}{{{\kappa_0}}}$ \cite{Chen2009Active,Bohnet2012A}            & \multirow{2}{*}{$0.038 \pm 0.002$}           & $T=120^{\circ}$C \\ 
					& 0.039            &     &0.040                    \\ \hline
					\multirow{2}{*}{Anti-resonant} &$   - {\left( {\frac{\pi }{{2\mathcal{F}}}} \right)^2}\frac{{{\Gamma}}}{{{\kappa_0}}}$           &  \multirow{2}{*}{$-0.019 \pm 0.002$}           & $T=120^{\circ}$C \\ 
					&$-$0.012   &  & $-$0.019   \\ 
					\hline \hline
			\end{tabular}}
			
			\begin{tablenotes}
				\footnotesize
				\item[*] In this work, $\mathcal{F}=3.07$; ${\Gamma} = 2\pi  \times 10.04 $ MHz; $\kappa_0=2\pi  \times 257$ MHz.
			\end{tablenotes}
			
		\end{threeparttable}
	\end{table*}

	{\bf Laser Linewidth.} Theoretically, analogous to the inhibited spontaneous emission, the linewidth of the inhibited laser has the potential to be narrower than that of the traditional resonant laser. Here, considering the cavity-modification effect, as well as homogeneous and inhomogeneous broadening, we provide the general expression of the laser linewidth as follows:
	\begin{align}\label{Eq4}
	\Delta {\nu _{\rm{L}}} 
	=&\frac{{{\Gamma ^2}}}{{4\pi n{\kappa _0}}}{N_{{\rm{sp}}}}\left[ {\frac{1}{{{\xi ^2}}} + \frac{{\left( {1 - \xi } \right){\Gamma _{\rm{g}}} + 2\left( {1 + \xi } \right){\Gamma _{\rm{e}}}}}{{4{\xi ^2}\left( {{\Gamma _{\rm{e}}} + {\Gamma _{\rm{g}}}} \right)}}\frac{n}{{{n_{\rm{s}}}}}} \right] \notag 
	\\
	&\frac{1}{{1 + {{\left( {\frac{{2\mathcal{F}}}{{\rm{\pi }}}} \right)}^2}{{\sin }^2}\left( {\frac{{\omega L}}{{\rm{c}}}} \right)}},
	\end{align}
	where ${N_{{\rm{sp}}}} = \frac{{{N_{\rm{e}}}}}{{{N_{\rm{e}}} - {N_{\rm{g}}}}}$ is the spontaneous-emission factor; $N_{\rm{e}}$ and $N_{\rm{g}}$ represent the populations of the excited and ground states, respectively; ${{\Gamma _{{\rm{e}}}}}$, ${{\Gamma _{{\rm{g}}}}}$, and ${{\Gamma _{{\rm{eg}}}}}$ are the decay rates of the atomic populations, and polarization; the coefficient $\xi  $ represents the inhomogeneous and homogeneous broadening; and ${n_{\rm{s}}} = \frac{{{\Gamma _{{\rm{eg}}}}}}{{2{g^2}}}\frac{{{\Gamma _{\rm{e}}}{\Gamma _{\rm{g}}}}}{{{\Gamma _{\rm{e}}} + {\Gamma _{\rm{g}}}}}$ is the homogeneous saturation intensity in units of number of photons. 
	
	Equation (\ref{Eq4}) shows four extra features compared with the classical Schawlow-Townes equation \cite{Schawlow1958Infrared}: (i) the first term represents the bad-cavity effect \cite{Kuppens1994Quantum} leading to linewidth narrowing, (ii) ${N_{{\rm{sp}}}}$ causes the linewidth broadening due to the incomplete inversion, (iii) the factors in square brackets is induced by the Doppler broadening and power broadening \cite{1996PRAkHOURY}, and (iv) the last term depicts the cavity-induced modification \cite{Heinzen1987Enhanced} following the absorption lineshape with the change of $\Delta \phi$.
	
	 It should be noted that the second term inside the square brackets is much bigger than the first one under the conditions of this work. Therefore, Eq. (\ref{Eq4}) can be further reduced, and the result shows that the expression of the laser linewidth is independent of the output power. Details about the laser linewidth are given in the Methods Section V. According to the last term of Eq. (\ref{Eq4}), the laser linewidth is expected to be narrowed by a factor of $\frac{1}{{1 + {{\left( {{{2{\cal F}} \mathord{\left/ {\vphantom {{2F} \pi }} \right. \kern-\nulldelimiterspace} \pi }} \right)}^2}}}$ for the inhibited laser compared with the resonant one, and this can be simplified to the classical bad-cavity expression \cite{1996PRAkHOURY} under the resonant condition. From Eq. (\ref{Eq4}), the quantum-limited linewidths of the resonant and inhibited lasers are 150 and 43 Hz, respectively (see Fig. \ref{ExpFig5}). Although the linewidth of the inhibited laser is affected by Doppler broadening, this thermal-atom scheme has the advantages of a compact structure and ease of operation, and it can solve the problem of pulse operations in the cold-atom scheme.  Moreover, if the Doppler broadening is suppressed by using cold atoms as the gain medium or the atomic-beam scheme \cite{Chen2009Active,Kazakov2014EFTF,PRL2020Holland,PRA2021Holland}, the quantum-limited linewidth of the inhibited laser can be further narrowed to the Hz level (details are given in Methods Section VI). In particular, a resonant continuous-wave superradiant laser with a linewidth of 40 mHz based on a hot atomic beam traversing an optical cavity has been proposed in Ref. \cite{PRL2020Holland}. If the gain in Eq. (\ref{Eq2}) can be improved by increasing the effective atomic number, so that meet the condition of laser oscillation, we expect to realize the inhibited laser by using the atomic-beam scheme, which is a feasible solution to achieve continuous-wave operation.

\begin{figure}[htbp]
		\centering
		\includegraphics[width=0.7\linewidth]{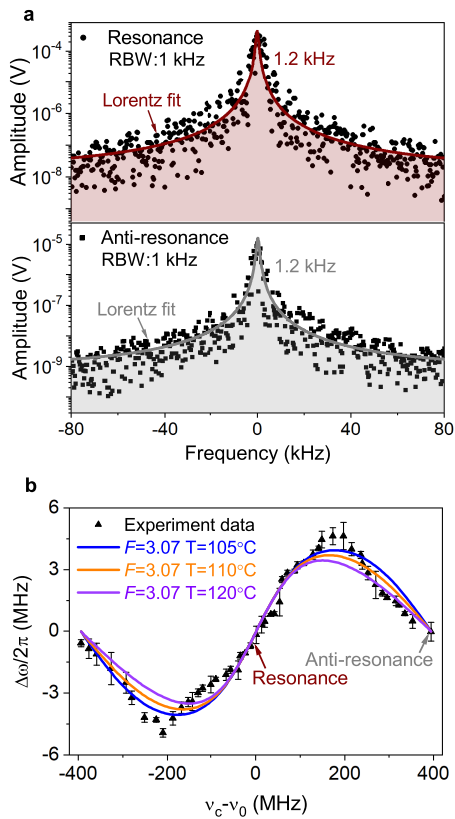}
		\caption{{\bf Linewidth and cavity-pulling characteristics.} {\bf a,} Beat-note spectra between the reference laser and the tested laser. Their resolution bandwidths were both 1 kHz with a sweep time of 362 ms. The beating spectrum (black circles) between the reference laser and the resonant tested laser was fitted by a Lorentzian function with a fitted linewidth of 1.2 kHz (red line). The Lorentz fitting linewidth of the beat-note spectrum (black squares) between the reference laser and the inhibited laser was also 1.2 kHz (grey line). {\bf b,} Frequency shift of the laser oscillation $\Delta$ as a function of cavity-frequency detuning from the atomic transition frequency ${\omega _{\rm{c}}}-\omega_0$, whose adjustable range was one FSR. We modify Eq. (\ref{Eq5}) with the laser rate equation to describe $\Delta$. The simulated results of $\Delta$ at different temperatures for $\mathcal{F}=3.07$ are shown as the solid lines, and the black triangles represent the experimental results. Error bars denote standard deviations from three measurements. The difference between the simulated data and experimental results is explained further in the main text.}
		\label{Fig3}
	\end{figure}

	We measured the laser linewidth by beating the tested laser against the reference laser, where the experimental setup is given in Fig. \ref{ExpFig1} in the Methods section. The cavity frequency of the reference laser coincided with the atomic transition frequency, while that of the tested laser was tunable by changing the cavity length. Limited to the intensity sensitivity of the photodetector, the cavity frequency of the reference laser should be coincident with the atomic resonance to improve the light intensity for optical heterodyning. It is difficult to measure the beat-note spectrum between two inhibited lasers due to their weak laser powers. Figure \ref {Fig3}a shows the beat-note spectra of the reference laser and the tested laser, in which the tested laser worked in the resonant and anti-resonant cavities, respectively. The beating linewidths were both 1.2 kHz. Although the experimental results cannot truly reflect the linewidth of the inhibited laser, it can show that the linewidth of the inhibited laser will not be wider than that of the resonant laser. In addition, limited by the technical noise, such as the cavity-length change induced by the temperature fluctuations of thermal atoms, the power fluctuations of the pumping laser, and the change of external magnetic field, the measured laser linewidth was wider than its corresponding quantum-limited linewidth. The influence of technical noise on the linewidth broadening is analyzed in detail in Methods Section VII.

{\bf Cavity-pulling characteristic.} As shown in Fig. \ref{Fig1}, the inhibited laser works in the flat anti-resonant regime, which is the center of two adjacent cavity modes. Therefore, the inhibited laser has the advantage of an enhanced suppression of the cavity-pulling effect. The relationship between the frequency shift of the oscillation frequency, i.e., $\Delta$, and the cavity-frequency detuning from the atomic transition ${\omega _{\rm{c}}}-\omega_{0}$, is analyzed comprehensively for spontaneous emission \cite{Daniel1988Radiative}, which is written as
	\begin{equation}\label{Eq5}
		\Delta {\rm{ = }}\frac{\Gamma }{{\rm{4}}}\frac{{{{\left( {\frac{{2\mathcal{F}}}{{\rm{\pi }}}} \right)}^2}\sin \left( {{{2\omega L} \mathord{\left/
							{\vphantom {{2\omega L} {\rm{c}}}} \right.
							\kern-\nulldelimiterspace} {\rm{c}}}} \right)}}{{1 + {{\left( {\frac{{2\mathcal{F}}}{{\rm{\pi }}}} \right)}^2}{{\sin }^2}\left( {{{\omega L} \mathord{\left/
							{\vphantom {{\omega L} {\rm{c}}}} \right.
							\kern-\nulldelimiterspace} {\rm{c}}}} \right)}}.
	\end{equation} 
	Therefore, for spontaneous radiation, the frequency shift caused by the cavity-frequency detuning is eliminated, not only when the atomic resonance coincides with one of the cavity resonances but also when the atomic resonance is halfway between two adjacent cavity resonances.

	 From Eq. (\ref{Eq5}), the cavity-pulling coefficients are equal to $\frac{{2\mathcal{F}}}{\pi }\frac{{{\Gamma }}}{{{\kappa_0}}}$ and ${\text{ - }}\frac{{2\mathcal{F}}}{\pi }\frac{{{\Gamma }}}{{{\kappa_0}}}\frac{{\text{1}}}{{{\text{1 + }}{{\left( {{{{\text{2}}\mathcal{F}} \mathord{\left/
								{\vphantom {{{\text{2}}\mathcal{F}} \pi }} \right.
								\kern-\nulldelimiterspace} \pi }} \right)}^{\text{2}}}}}$ utilizing $\mathcal{F} = \frac{{{\rm{\pi c}}}}{{L{\kappa _0}}}$, when the cavity is resonant and anti-resonant, respectively. The ratio between the two coefficients is approximately equal to $ - {\left( {\frac{{2\mathcal{F}}}{\pi }} \right)^2}$. Analogous to the spontaneous radiation, the ratio between the cavity-pulling coefficients when the cavity is resonant and anti-resonant is $ - {\left( {\frac{{2\mathcal{F}}}{\pi }} \right)^2}$ for the stimulated emission. The difference is that the pulling coefficient is $\frac{{{\Gamma}}}{{{\kappa_0}}}$ for the resonant bad-cavity laser \cite{Chen2009Active,Bohnet2012A}. Accordingly, the cavity-pulling coefficient of the inhibited laser is around $ - {\left( {\frac{\pi }{{2\mathcal{F}}}} \right)^2}\frac{{{\Gamma }}}{{{\kappa_0}}}$.

	More specifically, for the stimulated emission, we should consider the atom-cavity interactions. Therefore, Eq. (\ref{Eq5}) is further modified by the laser rate equation to obtain the frequency shift of the stimulated emission. The fitted results are depicted by solid lines in Fig. \ref{Fig3}b. In addition, we measured the frequency shift as a function of the cavity-frequency detuning from the atomic transition frequency. The experimental results (black triangles) were consistent with the fitted results. For comparison, the cavity-pulling coefficients discussed above are illustrated in Table \ref{table1}. Compared with the resonant condition, the suppression of the cavity-pulling effect of inhibited lasers was enhanced from 26 to 53 times.

	The deviation between the experimental and simulated results of the cavity-pulling characteristics in Fig. \ref{Fig3}b is analyzed. First, the low-reflectivity mirrors result in a significant loss in the cavity. Therefore, the initial measurement error may cause a deviation between the experimental and theoretical values of the finesse, since the intracavity multiple round-trip propagation of light will lead to error accumulation. Second, the piezoelectric ceramic (PZT) was used to tune the cavity length, i.e., the cavity-mode frequency. Ideally, the length change of the PZT should be linear with the voltage applied to the PZT. However, due to unavoidable manufacturing errors, this relationship is not perfectly linear, and leads to the deviation of cavity-frequency detuning between the experimental and simulated results. Third, the simulated result was obtained under the single-mode operation. In our experiment, although the higher-order transverse mode was suppressed, the environmental noise caused a the geometry change of the cavity, which would destroy the perfect single-mode operation to some extent. In summary, these factors will lead to the deviations between the experimental and simulated results, as shown in Fig. \ref{Fig3}b.

\section*{Conclusions}
\label{Conclusion}

In this work, we experimentally demonstrated an inhibited laser. It lased exactly at the center of two adjacent cavity modes, which is different from traditional resonant lasers. Compared with resonant superradiant lasers \cite{Chen2009Active,Bohnet2012A,PhysRevA.87.013821,PhysRevLett.123.103601,Stefan2020PhysRevA,PRL2020Holland}, the inhibited laser is inherently insensitive to cavity-length fluctuations, characterized by a further enhanced suppression on the cavity-pulling effect. The Doppler broadening can be further inhibited by the cold-atom or atomic-beam scheme \cite{Kazakov2014EFTF,PRL2020Holland,PRA2021Holland}. In the future, using the cold-atom as gain medium, we expect to achieve a sub-Hz-level linewidth inhibited laser with a cavity-pulling coefficient on the order $10^{-5}$ and an output power at the $\mu$W level by selecting an atomic transition with a lower decay rate, such as the Cs 1359 nm transition, to realize lasing. This result is comparable to the performances of resonant superradiant lasers using the natural linewidth at only the mHz-level as the transition level but solves its problem of a low output power. We expect to realize an anti-resonant superradiant AOC using the principle of the inhibited laser, which will greatly facilitate precision measurements for fundamental science, such as tests of the variations of fundamental constants, the gravitational potential of Earth, and the search for dark matter.

	\section*{Methods}
	\label{Method}
	
	{\bf I. Experimental details.} To acquire sufficient gain, we take advantage of the multilevel structure of the Cs atom and multiple atoms interacting with a single mode of an optical cavity.  As depicted in Fig. \ref{Fig1}, a cloud of thermal Cs atoms collected in the low-finesse F-P cavity were pumped by the 459 nm continuous-wave laser. For typical lasers, the cavity length is exactly equal to an integral multiple of the half-wavelength, i.e., the cavity-mode frequency resonates with the peak of the emission line for an atomic transition. However, the cavity length is equal to an odd multiple of the quarter wavelength, namely, the atomic transition frequency is halfway between two adjacent cavity modes, for the inhibited laser. In this work, the cavity length was tunable through the PZT, of which the displacement range was more than one-half wavelength of the laser oscillation. Therefore, the detuning range of the cavity frequency was larger that one FSR$ \sim $789 MHz.
	
\begin{figure}[hbtp]
	\centering
	\includegraphics[width=1\linewidth]{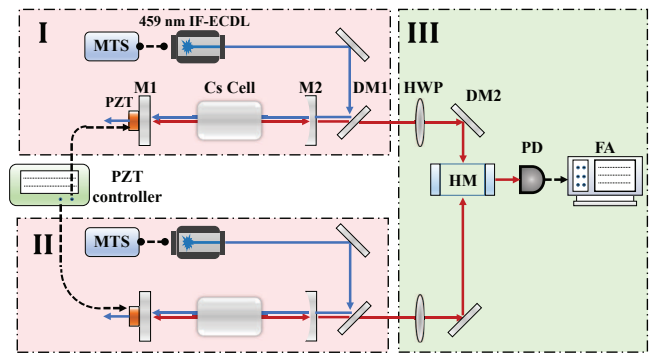}
	\caption{Experimental setup of the heterodyne measuring system between the reference laser and the inhibited laser. MTS, modulation transfer spectroscopy; IF-ECDL, interference filter configuration external cavity diode laser; PZT, piezoelectric ceramic; M1 and M2, cavity mirrors; DM1, dichroic mirror coated with anti-reflection coating at 1470 nm and high-reflection coating at 459 nm; HWP, halfwave plate; DM2, dichroic mirror coated with high-reflection at 1470 nm and anti-reflection at 459 nm; HM, 1470 nm heterodyne module; PD, photodetector; FA, frequency analyzer.} 
	\label{ExpFig0}
\end{figure}

	To measure the frequency shift of the laser oscillation with the beat-note method, we built another 1470 nm laser source as the frequency reference. The working schematic is shown in Fig. \ref{ExpFig0}, which consists of three modules: I. The tested-laser generating module, whose cavity length is adjustable over a half-wavelength transition at 1470 nm. II. The reference-laser generating module, whose cavity-mode frequency is tuned to be exactly equal to the central frequency of the gain medium. III. The heterodyne module, where the beat-note signals of the tested laser and the reference laser are recorded. In modules I and II, the 459 nm interference filter configuration external cavity diode laser (IF-ECDL), which was frequency stabilized by modulation transfer spectroscopy (MTS), pumped the Cs atoms inside the F-P cavity to realize the 1470 nm lasing. The displacement range of the PZT was adjustable in module I, while it was tuned to a fixed value in module II. When tuning the length of the F-P cavity in module I, we recorded the central frequency as well as the full width at half maximum (FWHM) of each beat-note spectrum by a frequency analyzer (FA).   
	
	{\bf II. Cavity-pulling coefficient.} The integrated Invar F-P cavity consisted of a plane mirror M$_1$ and a plane-concave mirror M$_2$ (radius of curvature $r=500$ mm) separated by a distance $L=190$ mm. Therefore, the mode sustained by the cavity had Gaussian transverse profiles, of which the spot radii on the cavity mirrors M$_1$ and M$_2$ were ${w_{s1}} = 0.429$ mm and ${w_{s2}} = 0.337$ mm, respectively. The equivalent mode volume was $V_{\rm{c}} = \frac{1}{4}L\pi {\left( {\frac{{{w_{s1}} + {w_{s2}}}}{2}} \right)^2} = 21.89$ mm$^3$. The cavity power decay rate was $\kappa_0 = 2\pi  \times 257$ MHz under resonant conditions, and the free spectral range was $\rm{FSR}=789$ MHz. Therefore, the cavity finesse was ${\mathcal{F}} = 3.07$ \cite{FS2006book}.

	The gain medium Cs atoms were pumped by the 459 nm laser through the velocity-selective mechanism. It was assumed that the pumping light intensity $I$ = 10 mW/mm$^2$, while the corresponding saturation light intensity ${I_{\rm{s}}} = \pi h{\rm{c}}\Gamma /3{\lambda ^3} = 1.27$ mW/cm${^2}$. Therefore, the saturation broadening of state $\left| 2 \right\rangle $ in Fig. \ref{Fig1} caused by the pumping laser was
	\begin{equation}\label{M1}
		\frac{{{\Gamma _{2}}}}{{2\pi }} = \frac{{{\Gamma _{21}} + {\Gamma _{23}} + {\Gamma _{24}}}}{{2\pi }}\sqrt {1 + {s}}=26.38 \,\rm{MHz} ,
	\end{equation}
	where ${\Gamma _{21}} = 0.793 \times {10^6}\,{{\rm{s}}^{ - 1}}$, ${\Gamma _{23}} = 3.52 \times {10^6}\,{{\rm{s}}^{ - 1}}$, and ${\Gamma _{24}} = 1.59 \times {10^6}\,{{\rm{s}}^{ - 1}}$ are the decay rates of the $\left| 2 \right\rangle  \to \left| 1 \right\rangle $, $\left| 2 \right\rangle  \to \left| 3 \right\rangle $, and $\left| 2 \right\rangle  \to \left| 4 \right\rangle $ transitions, respectively. $s$ is the saturation factor represented by $s = {I \mathord{\left/
			{\vphantom {I {{I_{\rm{s}}}}}} \right.
			\kern-\nulldelimiterspace} {{I_{\rm{s}}}}}$. According to the velocity-selective scheme, only atoms in the direction of the cavity mode with a velocity less than $\Delta \upsilon  = \frac{{{\Gamma _{2}}}}{{2\pi }} \times {\lambda _{21}}$ can be pumped to state $\left| 2 \right\rangle $ and then decay to state $\left| 3 \right\rangle $. Consequently, the Doppler broadening of $\left| 3 \right\rangle $ is $\frac{{{\Gamma _{\rm{D}}}}}{{2\pi }} = \Delta \upsilon /{\lambda _{34}}$. Since the spontaneous decay rate of the 1470 nm transition $\Gamma _0 = 2\pi  \times 1.81$ MHz \cite{JOSA1961Heavens}, the atomic decay rate was ${\Gamma} = \Gamma _0 + \Gamma _{\rm{D}} = 2\pi  \times 10.04$ MHz, which was much smaller than $\kappa$. Accordingly, the cavity-pulling coefficient in the resonant cavity is $P \approx {\Gamma  \mathord{\left/
			{\vphantom {\Gamma  \kappa }} \right.
			\kern-\nulldelimiterspace} \kappa_0 } = 0.039$.

	{\bf III. Intracavity photon number at steady state.} For the atomic number density $n' = 1.57 \times {10^{13}}$/cm$^3$ at a vapor-cell temperature of 100$^{\circ}$C \cite{webCsD}, the atomic number inside the cavity mode is 
	$N = \frac{1}{4}n'\pi {L_{\rm{cell}}}{\left( {\frac{{{w_{s1}} + {w_{s2}}}}{2}} \right)^2} = 1.81 \times {10^{11}} $, where the vapor-cell length is $L_{\rm{cell}}=10$ cm. Only the atoms with velocities between ${{ - \Delta \upsilon } \mathord{\left/
			{\vphantom {{ - \Delta \upsilon } {2 \sim }}} \right.
			\kern-\nulldelimiterspace} {2 \sim }}{{\Delta \upsilon } \mathord{\left/
			{\vphantom {{\Delta \upsilon } 2}} \right.
			\kern-\nulldelimiterspace} 2} $ can be pumped to the 7P$_{1/2}$ state. According to the Maxwell speed distribution, the effective atomic number $N_{\rm{eff}}$ is given by
	\begin{align}\label{M2}
		{N_{{\rm{eff}}}} &= N\int_{ - \frac{1}{2}\Delta \upsilon }^{\frac{1}{2}\Delta \upsilon } {\sqrt {\frac{m}{{2\pi {k_{\rm{B}}}T}}} } \exp \left( {\frac{{ - m{\upsilon ^2}}}{{2{k_{\rm{B}}}T}}} \right)d\upsilon  \notag\\
		&=5.71 \times {10^9} ,
	\end{align}
	where $m$ is the atomic mass, and ${{k_{\rm{B}}}}$ is the Boltzmann constant.

\begin{figure}[hbtp]
	\centering
	\includegraphics[width=1\linewidth]{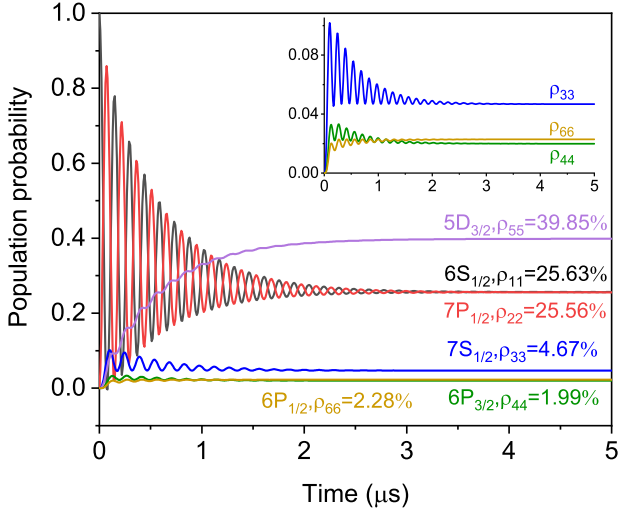}
	\caption{Numerical results of the population probability of each state ${\rho _{ii}}$ in Fig. \ref{Fig1}a, under $g = 1.99 \times {10^5}\,{\rm{s}^{ - 1}}$, $\Omega  = 4.30 \times {10^7}\,{\rm{s}^{ - 1}}$, and $N_{\rm{eff}} = 5.71 \times {10^{9}}$.} 
	\label{ExpFig1}
\end{figure}

	Utilizing the density matrix equations, the intracavity photon number at steady state as a function of the phase shift $\Delta \phi$ (or cavity-frequency detuning, ${\nu _c} - {\nu _0}$) is obtained. The atomic energy level is shown in Fig. \ref{Fig1}a, where the energy states are labelled as $\left| i \right\rangle $. Under the conditions of this work, the homogeneous intensity broadening was much larger than the inhomogeneous Doppler broadening, which is analyzed in detail in Methods Section V. Using the rotating wave approximation (RWA), the density matrix equations for Cs atoms interacting with the 459 nm pumping laser are expressed as follows: 
	\begin{align}\label{M3}
		&\frac{{d{\rho _{11}}}}{{dt}} =  - \Omega \rho _{{\text{12}}}^I + {\Gamma _{21}}{\rho _{22}} + {\Gamma _{41}}{\rho _{44}} + {\Gamma _{61}}{\rho _{66}}, \notag\\
		&\frac{{d{\rho _{22}}}}{{dt}} = \Omega \rho _{{\text{12}}}^I - \left( {{\Gamma _{21}} + {\Gamma _{23}} + {\Gamma _{25}}} \right){\rho _{22}}, \notag \\
		&\frac{{d{\rho _{33}}}}{{dt}} = {\Gamma _{23}}{\rho _{22}} - \left( {{\Gamma _{34}} + {\Gamma _{36}}} \right){\rho _{33}}- \frac{{{\rho _{33}} - {\rho _{44}}}}{{{\tau _{\text{cyc}}}}} \notag \\
		&{\left( {\frac{{2g\sqrt {n + 1} }}{{\sqrt {{\Delta ^2} + 4{g^2}\left( {n + 1} \right)} }}} \right)^2}{\sin ^2}\left( {\frac{{\sqrt {{\Delta ^2} + 4{g^2}\left( {n + 1} \right)} }}{2}{\tau _{{\text{cyc}}}}} \right), \notag \\
		&\frac{{d{\rho _{44}}}}{{dt}} = {\Gamma _{{\text{34}}}}{\rho _{{\text{33}}}} + {\Gamma _{{\text{54}}}}{\rho _{55}} - {\Gamma _{{\text{41}}}}{\rho _{{\text{44}}}}+\frac{{{\rho _{33}} - {\rho _{44}}}}{{{\tau _{\text{cyc}}}}}  \notag\\
		&{\left( {\frac{{2g\sqrt {n + 1} }}{{\sqrt {{\Delta ^2} + 4{g^2}\left( {n + 1} \right)} }}} \right)^2}{\sin ^2}\left( {\frac{{\sqrt {{\Delta ^2} + 4{g^2}\left( {n + 1} \right)} }}{2}{\tau _{{\text{cyc}}}}} \right), \notag \\
		&\frac{{d{\rho _{{\text{55}}}}}}{{dt}} = {\Gamma _{{\text{25}}}}{\rho _{{\text{22}}}}{\text{ - }}\left( {{\Gamma _{{\text{54}}}} + {\Gamma _{{\text{56}}}}} \right){\rho _{{\text{55}}}}, \notag \\
		&\frac{{d{\rho _{{\text{66}}}}}}{{dt}} = {\Gamma _{{\text{36}}}}{\rho _{{\text{33}}}}{\text{ + }}{\Gamma _{{\text{56}}}}{\rho _{{\text{55}}}} - {\Gamma _{{\text{61}}}}{\rho _{{\text{66}}}}, \notag \\
		&\frac{{d\rho _{{\text{12}}}^I}}{{dt}} = \frac{1}{2}\Omega \left( {{\rho _{11}} - {\rho _{{\text{12}}}}} \right) + \rho _{{\text{12}}}^R\Delta ' - \frac{1}{2}{\Gamma _{21}}\rho _{{\text{12}}}^I, \notag \\
		&\frac{{d\rho _{{\text{12}}}^R}}{{dt}} =  - \rho _{{\text{12}}}^I\Delta ' - \frac{1}{2}{\Gamma _{21}}\rho _{{\text{12}}}^R, \notag \\
		&\frac{{dn}}{{dt}} = {N_{\rm{eff}}}\frac{{{\rho _{33}} - {\rho _{44}}}}{{{\tau _{{\text{cyc}}}}}}{\left( {\frac{{2g\sqrt {n + 1} }}{{\sqrt {{\Delta ^2} + 4{g^2}\left( {n + 1} \right)} }}} \right)^2} \notag \\
		&{\sin ^2}\left( {\frac{{\sqrt {{\Delta ^2} + 4{g^2}\left( {n + 1} \right)} }}{2}{t _{{\text{int}}}}} \right) - \eta {\kappa_0 }n.
	\end{align}
	$\Omega$ is the Rabi frequency, and ${\Gamma _{ij}}$ represents the rate of decay from $\left| i \right\rangle $ to $\left| j \right\rangle $. ${\tau _{{\rm{cyc}}}} = \frac{1}{\Omega } + \frac{1}{{{\Gamma _{23}} + {\Gamma _{25}}}} + \frac{1}{{{\Gamma _{34}} + {\Gamma _{36}}}} + \frac{1}{{{\Gamma _{41}}}}=28.0$ $\mu {\rm{s}}$ is the cycle time for Cs atoms through a complete transition of $6{{\rm S}_{1/2}} \to 7{{\rm P}_{1/2}} \to 7{{\rm S}_{1/2}} \to 6{{\rm P}_{3/2}}$. The interaction time between the atoms and the cavity mode is given by ${t_{{\rm{int}}}} \approx \frac{1}{{{\Gamma _{34}} + {\Gamma _{36}} + {\Gamma _{41}}}}=19.8$ ns. Ideally, we would simplify the equations by setting the frequency detuning between the pumping laser and the atomic transition of $\left| 1 \right\rangle $ to $\left| 2 \right\rangle $ to be zero. $\rho _{{\text{12}}}^I$ and $\rho _{{\text{12}}}^R$ represent the energy shift and the power broadening, respectively. ${\rho _{ii}}$ denotes the population probability of atoms in the corresponding state, and the result is shown in Fig. \ref{ExpFig1}. $\Delta  = \omega  - {\omega _0}$ is the frequency detuning of the laser oscillation from the atomic transition. Since ${\Delta ^2} \ll 4{g^2}\left( {n + 1} \right)$, we assume that $\Delta  = 0$ in the main text. To verify the correctness of this assumption, we give the most accurate description of the detuning $\Delta$ in Eq. (\ref{M3}). The photon number at steady state is calculated by inserting 
	the fitted result of $\Delta$ in Fig. \ref{Fig3}b into Eq. (\ref{M3}).

\begin{figure}[hbtp]
	\centering
	\includegraphics[width=1\linewidth]{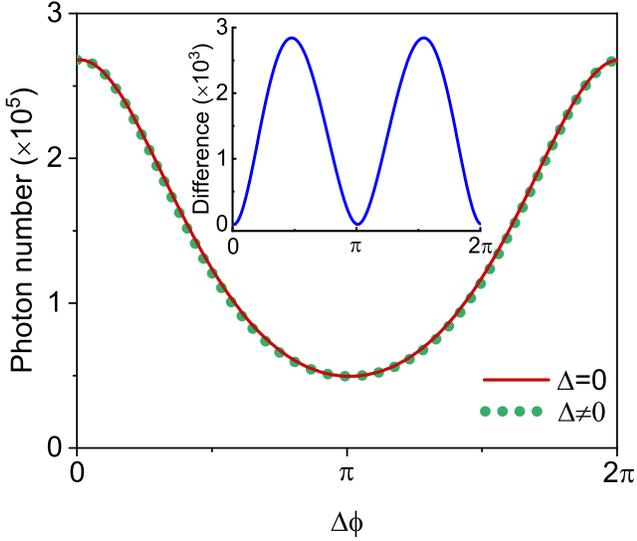}
	\caption{Photon number at steady state as a function of the phase shift for $g = 1.99 \times {10^5}\,{\rm{s}^{ - 1}}$, $\Omega  = 4.30 \times {10^7}\,{\rm{s}^{ - 1}}$, and $N_{\rm{eff}} = 5.71 \times {10^{9}}$. The red line and the green dotted line represent the photon number when  $\Delta=0$ and $\Delta  \ne {\text{0}}$, respectively. The inset shows the difference between the photon numbers when $\Delta=0$ and $\Delta  \ne {\text{0}}$.} 
	\label{ExpFig2}
\end{figure}

	The results of the intracavity photon number at steady state with and without considering the detuning $\Delta$ are shown as the green dotted line and the red solid line in Fig. \ref{ExpFig2}. This shows that there was little difference between the photon number obtained by Eq. (\ref{Eq2}) and the last equation of Eq. (\ref{M3}). The difference between photon numbers obtained by the two equations is shown in the inset of Fig. \ref{ExpFig2}, which illustrates that the differences are both zero when the cavity is
	resonant and anti-resonant. The difference is eliminated when the mode frequency exactly coincides with the center frequency of the gain profile. In addition, when the mode frequency is tuned to the center of two adjacent cavity resonances, the effects of cavity-pulling of the two adjacent cavity modes on the laser frequency are equal and opposite. Hence, the difference is also zero for the inhibited laser. This result demonstrates that the approximation used in the main text is reasonable.

\begin{figure}[hbtp] 
	\centering
	\includegraphics[width=1\linewidth]{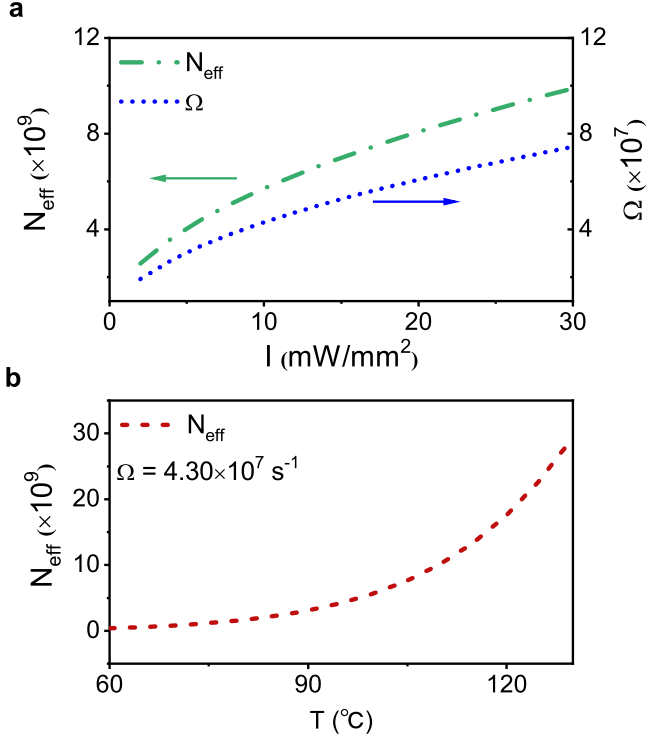}
	\caption{{\bf a,} Effective atomic number and Rabi frequency vs. the pumping light intensity at a vapor-cell temperature of 100$^{\circ}$C. {\bf b,} Effective atomic number as a function of the vapor-cell temperature under a pumping light intensity of 10 mW/mm$^2$.} 
	\label{ExpFig3}
\end{figure}

	To further characterize the output laser power as a function of the phase shift at different pumping efficiencies and atomic densities, the pumping light intensity and the vapor-cell temperature are adjustable, as depicted in Fig. \ref{Fig2}a. According to Eq. (\ref{M2}), the intracavity effective atomic number is influenced by both the pumping light intensity and the vapor-cell temperature, while the Rabi frequency of the pumping laser is relative to the pumping light intensity, which is described as $\Omega  = \sqrt {\frac{{3{\lambda _{21}}^3{\Gamma _{21}}I}}{{2\pi hc}}}$. $N_{\rm{eff}}$ and $\Omega$ as functions of $I$ and $N_{\rm{eff}}$ vs. $T$ are shown in Fig. \ref{ExpFig3}a and b, respectively.

\begin{figure}[hbtp]
	\centering
	\includegraphics[width=1\linewidth]{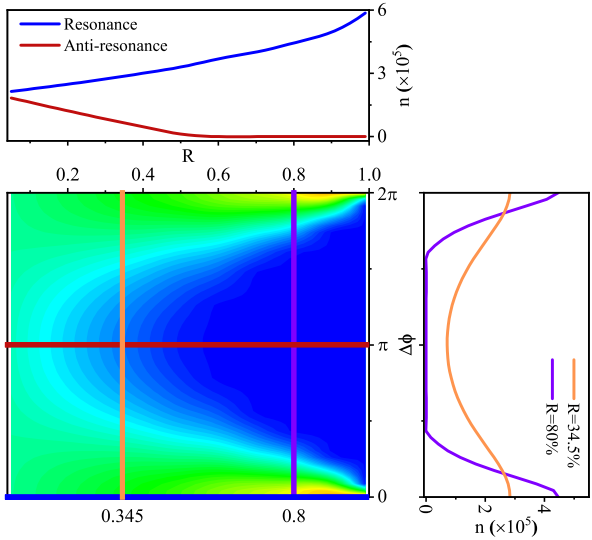}
	\caption{Intracavity photon number $n$ as a function of the phase shift $\Delta \phi $ at different cavity-mirror reflectivity $R$ values for $g = 1.99 \times {10^5}\,{\rm{s}^{ - 1}}$, $\Omega  = 4.30 \times {10^7}\,{\rm{s}^{ - 1}}$, and $N_{\rm{eff}} = 5.71 \times {10^{9}}$. The blue and red lines represent the photon number when the cavity is resonant and anti-resonant, respectively. The orange and the purple lines show the results at $R=34.5\%$ and $R=80\%$ ($n \le 1$), respectively.} 
	\label{ExpFig4}
\end{figure}

	{\bf IV. Photon number as function of cavity decay rate.} According to Eq. (\ref{Eq2}), the function of the intracavity photon number $n$ with phase shift $\Delta \phi $ varies with the cavity-mirror reflectivity $R$, namely, the cavity power loss rate $\kappa$. When the cavity is anti-resonant ($\Delta \phi  = \left( {2q + 1} \right)\pi $), the intracavity photon number decreased with the increase in the reflectivity, which is shown in Fig. \ref{ExpFig4}. The intracavity photon number for the inhibited laser was smaller than 1 when the reflectivity increased to 80$\%$ with $g = 1.99 \times {10^5}\,{\rm{s}^{ - 1}}$, $\Omega  = 4.30 \times {10^7}\,{\rm{s}^{ - 1}}$, and $N_{\rm{eff}} = 5.71 \times {10^{9}}$. Nevertheless, $n$ could be further improved with a higher pumping light intensity and a higher atomic number density.

	{\bf V. General expression of laser linewidth.} Considering the effects of inhomogeneous and homogeneous broadening and the cavity-induced modifications, this work provides a complete expression of the laser linewidth for the four-level atomic structure. First, we show the definitions of the parameters involved in the calculation as follows:
	
	(1) ${{\Gamma _{{\rm{e}}}}}$ and ${{\Gamma _{{\rm{g}}}}}$: decay rates of the atomic populations of the upper and lower levels, respectively.
	
	(2) ${{\Gamma _{{\rm{eg}}}}}$: decay rate of the atomic polarization. 
	
	(3) $\Delta {\omega _{\rm{D}}}$: Doppler broadening.
	
	(4) $\alpha  = \frac{{2\Delta {\omega _{\rm{D}}}}}{{{\Gamma _{{\rm{eg}}}}}}$: dimensionless inhomogeneous broadening width.
	
	(5) ${n_{\rm{s}}} = \frac{{{\Gamma _{{\rm{eg}}}}}}{{4{g^2}}}\frac{{{\Gamma _{\rm{e}}}{\Gamma _{\rm{g}}}}}{{{\Gamma _{\rm{e}}} + {\Gamma _{\rm{g}}}}}$: homogeneous saturation intensity in units of number of photons.
	
	(6) ${n}$: intensity of the laser light in units of number of photons.

	(7) $\beta  = \sqrt {1 + {n \mathord{\left/
				{\vphantom {n {{n_{\rm{s}}}}}} \right.
				\kern-\nulldelimiterspace} {{n_{\rm{s}}}}}}  $: dimensionless homogeneous intensity broadening width.

In this work, the thermal Cs atoms inside the optical cavity were pumped by the 459 nm laser through q velocity-selective mechanism. For the pumping light intensity of 10 mW/mm$^2$, only atoms in the direction of the cavity mode with a velocity of $\left| \vartheta  \right| \le {{\Delta \vartheta } \mathord{\left/
		{\vphantom {{\Delta \vartheta } 2}} \right.
		\kern-\nulldelimiterspace} 2} = 6.05$ m/s could be pumped to the excited state. Therefore, the Doppler broadening $\Delta {\omega _{\rm{D}}} = \frac{\vartheta }{{\rm{c}}}{\omega _0} = 25.9 \times {10^6}$ s$^{-1}$, where $\omega _0$ represents the transition frequency. Here, ${\Gamma _{{\rm{eg}}}} = 11.4 \times {10^6}$ s$^{-1}$, ${\Gamma _{{\rm{e}}}} = 17.6 \times {10^6}$ s$^{-1}$, ${\Gamma _{{\rm{g}}}} = 32.4 \times {10^6}$ s$^{-1}$. Thus, $\alpha  = {{{\text{2}}\Delta {\omega _{\text{D}}}} \mathord{\left/
		{\vphantom {{{\text{2}}\Delta {\omega _{\text{D}}}} {{\Gamma _{{\text{eg}}}}}}} \right.
		\kern-\nulldelimiterspace} {{\Gamma _{{\text{eg}}}}}} = 4.54$.

	The intracavity photon number at steady state is influenced by the detuning of cavity-mode frequency from atomic transition. In this work, the intracavity photon number $n$ was in the range of $5.51 \times {10^4}$ to $2.70 \times {10^5}$ under the conditions of $g = 1.99 \times {10^5}\,{\rm{s}^{ - 1}}$, $\Omega  = 4.30 \times {10^7}\,{\rm{s}^{ - 1}}$, and $N_{\rm{eff}} = 5.71 \times {10^{9}}$. ${{n_{\rm{s}}}}=819$ with the atom-cavity coupling constant $g = 1.99 \times {10^5}$ s$^{-1}$. Therefore, $\beta$ was in the range of $8.27$ to $18.18$. Consequently, ${\alpha  \mathord{\left/
			{\vphantom {\alpha  \beta }} \right.
			\kern-\nulldelimiterspace} \beta } < 1$, and the intensity broadening was slightly larger than the Doppler broadening. 
	
	Moreover, the waist radius ${w_0} = \frac{{{w_{s1}} + {w_{s2}}}}{2} = 0.383$ mm, and the atomic velocity $\left| \vartheta  \right| \le 6.05$ m/s. Therefore, the atomic transit time $\tau  \ge \frac{{2{w_0}}}{{\left| \vartheta  \right|}} = 0.127$ ms, which is much longer than the atomic transition lifetime ${\Gamma _{{\rm{eg}}}}^{{\rm{ - 1}}}$ \cite{PRA2008Yu}. Finally, we considered the influence of both inhomogeneous broadening and the cavity-modified effect, giving a general expression of the laser linewidth $\Delta {\nu _{\rm{L}}}$ under the assumption of $\tau  \gg {\Gamma _{{\rm{eg}}}}^{{\rm{ - 1}}}$, as follows:
\begin{align}\label{M4}
&\Delta {\nu _{\rm{L}}} = \notag 
\\
& \frac{\kappa }{{4\pi n}}{\left( {\frac{{\Gamma '}}{{\Gamma ' + \kappa }}} \right)^2}{N_{{\rm{sp}}}}\left[ {1 + \frac{{\left( {1 - \xi } \right){\Gamma _{\rm{g}}} + 2\left( {1 + \xi } \right){\Gamma _{\rm{e}}}}}{{4\left( {{\Gamma _{\rm{e}}} + {\Gamma _{\rm{g}}}} \right)}}\frac{n}{{{n_{\rm{s}}}}}} \right],
\end{align}	
where $\kappa  = \eta {\kappa _0}$, $\kappa _0$ represents the cavity dissipation rate under the resonant condition. $\eta  = 1 + {\left( {\frac{{2\mathcal{F}}}{{\rm{\pi }}}} \right)^2}{\sin ^2}\left( {\frac{{\omega L}}{{\rm{c}}}} \right) \ge 1$ is the loss coefficient, which is depicted in the main text. Here, we have let $\Gamma ' = {\Gamma  \mathord{\left/
		{\vphantom {\Gamma  \xi }} \right.
		\kern-\nulldelimiterspace} \xi }$ to make the expression of Eq. (\ref{M4}) more similar to the one depicting the homogeneous broadening case \cite{PRA1993Kolobov}. The coefficient $\xi  $ describes the inhomogeneous and homogeneous broadening, as
\begin{align}\label{M5}
\xi   = \frac{2}{{\sqrt {\rm{\pi }} }}\frac{\beta }{\alpha }\frac{{\exp \left( { - {{{\beta ^2}} \mathord{\left/
					{\vphantom {{{\beta ^2}} {{\alpha ^2}}}} \right.
					\kern-\nulldelimiterspace} {{\alpha ^2}}}} \right)}}{{{\rm{erfc}}\left( {{\beta  \mathord{\left/
					{\vphantom {\beta  \alpha }} \right.
					\kern-\nulldelimiterspace} \alpha }} \right)}} - \frac{{2{\beta ^2}}}{{{\alpha ^2}}},\notag 
				\\
	{\rm{erfc}}\left( z \right) = 1 - \frac{2}{{\sqrt {\rm{\pi }} }}\int_0^z {dt\exp \left( { - {t^2}} \right)}. 
\end{align}
${N_{{\rm{sp}}}} = \frac{{{N_{\rm{e}}}}}{{{N_{\rm{e}}} - {N_{\rm{g}}}}}$ is the spontaneous-emission factor, $N_{\rm{e}}$ and $N_{\rm{g}}$ represent the populations of the excited and ground states, respectively. 

\begin{figure}[hbtp]
	\centering
	\includegraphics[width=1\linewidth]{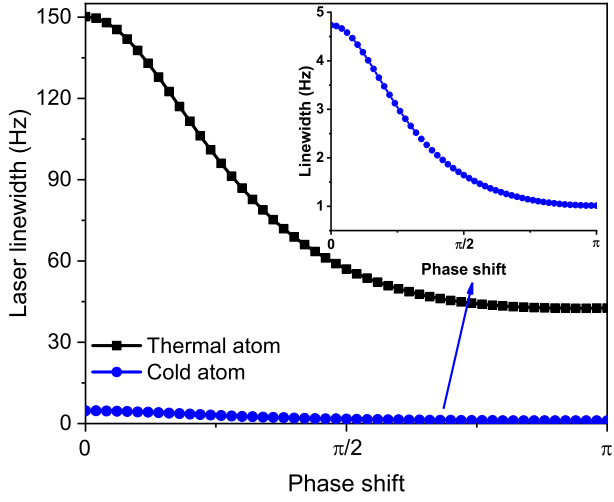}
	\caption{Quantum-limited linewidth $\Delta {\nu _{\rm{L}}}$ as a function of the phase shift $\Delta \phi $. The black-square dotted line and the blue dotted line represent the results of thermal- and cold-atom as gain medium, respectively. } 
	\label{ExpFig5}
\end{figure}

Furthermore, in the bad-cavity limit, $\Gamma ' \ll \kappa $. Therefore, we replace ${\left( {\frac{{\Gamma '}}{{\Gamma ' + \kappa }}} \right)^2}$ with ${\left( {\frac{{\Gamma '}}{\kappa }} \right)^2}$ and simplify Eq. (\ref{M4}) as
\begin{align}\label{M6}
\Delta {\nu _{\rm{L}}} 
=&\frac{{{\Gamma ^2}}}{{4\pi n{\kappa _0}}}{N_{{\rm{sp}}}}\left[ {\frac{1}{{{\xi ^2}}} + \frac{{\left( {1 - \xi } \right){\Gamma _{\rm{g}}} + 2\left( {1 + \xi } \right){\Gamma _{\rm{e}}}}}{{4{\xi ^2}\left( {{\Gamma _{\rm{e}}} + {\Gamma _{\rm{g}}}} \right)}}\frac{n}{{{n_{\rm{s}}}}}} \right] \notag 
\\
&\frac{1}{{1 + {{\left( {\frac{{2\mathcal{F}}}{{\rm{\pi }}}} \right)}^2}{{\sin }^2}\left( {\frac{{\omega L}}{{\rm{c}}}} \right)}}.
\end{align}
The analysis of the features related to the laser linewidth is given in the main text. It should be noted that the second term inside the square brackets is much larger than 1 under the conditions of this work. Then, Eq. (\ref{M6}) can be reduced as follows:
\begin{align}\label{M7}
\Delta {\nu _{\text{L}}} = &\frac{{{\Gamma ^2}}}{{4\pi {\kappa _0}}}{N_{{\text{sp}}}}\left[ {\frac{{\left( {1 - \xi } \right){\Gamma _{\text{g}}} + 2\left( {1 + \xi } \right){\Gamma _{\text{e}}}}}{{4{\xi ^2}\left( {{\Gamma _{\text{e}}} + {\Gamma _{\text{g}}}} \right)}}\frac{{\text{1}}}{{{n_{\text{s}}}}}} \right] \notag 
\\
&
\frac{1}{{1 + {{\left( {\frac{{2\mathcal{F}}}{\pi }} \right)}^2}{{\sin }^2}\left( {\frac{{\omega L}}{{\text{c}}}} \right)}}.
\end{align}
Accordingly, the expression of the linewidth is independent of the output power. Also, we calculated the quantum-limited linewidth with the change of the phase shift of the cavity, as shown in the black-square dotted line of Fig. \ref{ExpFig5}. The results agree with both Eq. (\ref{M6}) and (\ref{M7}). This indicates that the limit linewidths were 150 and 43 Hz when the cavity was resonant and anti-resonant, respectively. This result is limited by the Doppler broadening in the thermal atomic system. To explore the narrower quantum-limited linewidth, next, we used the cold-atom and atomic-beam methods to reduce the Doppler broadening and separately calculated the laser linewidth. 
	
{\bf VI. Laser linewidths using cold-atom and atomic-beam schemes}. First, we analyzed the cold-atom scheme. Assuming that the Cs atoms inside the optical cavity were cooled and trapped by the magneto-optical trap (MOT) at 894 nm. Meanwhile, the 459 nm laser pumped the cold Cs atoms to realize the 1470 nm laser. This design was used to reduce the light shift induced by the cooling and pumping light, because the energy levels of the 1470 nm transition and the cooling and pumping light are separated.

Here, we assumed that the Cs atoms, which were confined in an optical cavity with the finesse and dissipation rate being same as in the thermal-atom conditions, could be cooled to a temperature of 200 $\mu$K using a two-dimensional MOT \cite{PRL2019CJF}. The effective atomic number could also reach ${N_{{\rm{eff}}}} = 5.71 \times {10^9}$. The only difference was that the atomic decay rate $\Gamma  \approx {\Gamma _{{\rm{eg}}}}$, because the Doppler broadening was much smaller than the natural linewidth in the cold-atom system. Under this condition, the Doppler broadening $\Delta {\omega _{\rm{D}}} = \omega \sqrt {\frac{{2{k_{\rm{B}}}T}}{{m{{\rm{c}}^2}}}}  = 2{\rm{\pi }} \times 0.1$ MHz. Therefore, the ratio of the Doppler broadened inhomogeneous width to the power broadened homogeneous width $\frac{\alpha }{\beta } = \frac{{2\Delta {\omega _{\rm{D}}}}}{{{\Gamma _{{\rm{eg}}}}\left( {1 + {n \mathord{\left/
					{\vphantom {n {{n_{\rm{s}}}}}} \right.
					\kern-\nulldelimiterspace} {{n_{\rm{s}}}}}} \right)}} \ll 1$. The main contribution of the linewidth broadening comes from power broadening. The laser works in the homogeneous limit, and its linewidth can be reduced to a simple expression as follows:
\begin{align}\label{M8}
\Delta {\nu _{\rm{L}}} &= \frac{\kappa }{{{\rm{4\pi }}n}}{N_{\rm{sp}}}{\left( {\frac{\Gamma }{{\Gamma  + \kappa }}} \right)^2}\left( {1 + \frac{{{\Gamma _{\rm{e}}}}}{{{\Gamma _{\rm{e}}} + {\Gamma _{\rm{g}}}}}\frac{n}{{{n_{\rm{s}}}}}} \right)
\notag\\
& = \frac{{{\Gamma ^2}}}{{4\pi n{\kappa _0}}}{N_{{\rm{sp}}}}\left( {1 + \frac{{{\Gamma _{\rm{e}}}}}{{{\Gamma _{\rm{e}}} + {\Gamma _{\rm{g}}}}}\frac{n}{{{n_{\rm{s}}}}}} \right)\frac{1}{{1 + {{\left( {\frac{{2\mathcal{F}}}{{\rm{\pi }}}} \right)}^2}{{\sin }^2}\left( {\frac{{\omega L}}{{\rm{c}}}} \right)}}. 
\end{align}
Compared with Eq. (\ref{M6}), the difference of Eq. (\ref{M8}) comes from the fourth term, which only reflects the power broadening rather than the inhomogeneous broadening. The result is depicted by the blue dotted line in Fig. \ref{ExpFig5}. It shows that the quantum-limited linewidth was narrowed from 4.74 to 1.01 Hz when the cavity was tuned from resonant to anti-resonant. Moreover, the limited linewidth could be further narrowed to the sub-Hz level if we increased the cavity-dissipation rate or selected an atomic transition with a lower decay rate, such as the 1359 nm transition, as the inhibited laser. In addition, utilizing cold atoms as the gain medium, the cavity-pulling coefficient of the inhibited laser could be significantly reduced to the order of $10^{-5}$. Thus, the sensitivity of the laser frequency to the cavity frequency fluctuations was strongly suppressed. 

 Another approach to suppress the Doppler broadening is to couple a beam of moving atomic dipoles to a single mode of bad-cavity \cite{Chen2009Active,Kazakov2014EFTF,PRL2020Holland,PRA2021Holland}. It has been proposed that the 40-mHz-linewidth resonant suppradiant laser can be realized by using the atomic-beam scheme \cite{PRL2020Holland}. If the gain in laser rate equation can be improved by increasing the effective atomic number, so that meet the condition of laser oscillation, it is expected to realize the inhibited laser based on the atomic-beam scheme. In this case, $^{40}$Ca atoms are evaporated to form an atomic beam and effuse from the oven. After emerging from the oven, the atoms are precooled and prepared in the excited state before entering the anti-resonant cavity. The cavity is aligned transverse to the atomic beam to suppress the motional and collisional effects. The residual cavity thermal noise can be effectively restrained by placing the cavity inside a vacuum chamber with an actively controlled vibration isolation table combined with an additional thermally isolated passive heat shield \cite{Kessler2012NP}.   
	
Utilizing the atomic-beam system, the inhomogeneous Doppler broadening could reach around 100 kHz, which is much smaller than the intensity broadening. Therefore, the laser operates within the homogeneous limit, and the quantum-limited linewidth is also expressed by Eq. (\ref{M8}). The linewidth of the inhibited laser can be narrowed to the Hz level, and the problem of pulsed operation in the cold-atom scheme can be solved.

\textbf{VII. Linewidth broadening induced by technical noises.} In this work, a cloud of Cs atoms was pumped by the 459 nm laser through the velocity-selective mechanism. To increase the effective atomic number, the atoms inside the vapor cell were heated to around 100 $^\circ$C. Due to the present experimental scheme, there are mainly three kinds of technique noise that broaden the laser linewidth. The analysis of each type of  technical noise is given below.
	
	\textbf{a. Temperature fluctuations of atomic vapor cell.}

	First, assuming that the cavity-length is free running, we measured the frequency shift of the 1470 nm laser vs. the vapor-cell temperature, which has been reported in Ref. \cite{Pan2018IEEE}. The results show that the frequency of the laser changed almost linearly with the atomic temperature, with a slope of 692 kHz/$^\circ$C. The Allan deviation of the vapor-cell temperature was better than $1 \times {\rm{1}}{{\rm{0}}^{{\rm{ - 5}}}}$ at a temperature of 100$^\circ$C on a short time scale. Therefore, the linewidth broadening caused by the temperature fluctuations of the vapor cell was around 692 Hz.
	
	Superficially, the frequency shift caused by temperature fluctuations is serious. However, by introducing a reference light for the same measurement, it was found that the frequency shift was actually introduced due to the change of the equivalent cavity length caused by the cell ends' glass wall thermal expansion. The linewidth broadening did not indicate a collision broadening of 692 Hz, because this broadening could be greatly suppressed by eliminating the influence of cavity-length fluctuations, as shown in the analysis below.
	
	To analyze the impact of collision broadening on the laser linewidth caused by temperature fluctuations individually, we stabilized the cavity length by the optical phase loop locking (OPLL) technique in the system of the dual-wavelength AOC, whose experimental details are shown in Ref. \cite{Shi2019OE}. After cavity-length stabilization, the slope of the frequency shift of the 1470 nm laser with the change of the temperature fluctuations was below 45 $\pm$ 1.2 kHz/$^\circ$C. Also, the stability of the vapor-cell temperature was $1 \times {\rm{1}}{{\rm{0}}^{{\rm{ - 5}}}}$ at a temperature of 100$^\circ$C on the short time scale. This indicated that the collision broadening caused by the atomic temperature fluctuations was below 45 Hz. Therefore, the impact of collision broadening on the short-term stability of the 1470 nm laser was small. However, on the long timescales, the effect of the collision broadening on the long-term relative frequency stability of the 1470 nm laser is indeed an issue that needs to be solved. 
	
	In this work, the cavity-length was free running. Therefore, the linewidth broadening caused by the temperature fluctuations of the atomic vapor cell was around 700 Hz, where the collision broadening was about 45 Hz. To further reduce the influence of temperature fluctuations on the laser linewidth, the cavity will be placed into a vacuum chamber and a cold-atom system will be built in our future work.

	\textbf{b. Power fluctuations of 459 nm pumping laser.} 
	
	The linewidth of the 1470 nm laser was also influenced by the power stability of the pumping laser. To further evaluate the influence of the power stability of the pumping laser on the laser linewidth, we measured the laser frequency variations with the power of the pumping laser, as shown in Ref. \cite{Shi2019OE}. An approximately linear relationship was obtained, with a slope of $-36.1 \pm 0.94$ kHz/mW. The Allan deviation of the 459 nm laser power was ${\rm{1}}{\rm{.28}} \times {\rm{1}}{{\rm{0}}^{{\rm{ - 4}}}}$ at 1 s, and the pumping power was about 10 mW, which means that the linewidth broadening caused by the power fluctuations of the pumping laser was around 45 Hz. With further power stabilization of the pumping laser, we expect a linewidth broadening of the 1470 nm laser because the power fluctuations of the pumping laser can be reduced to below the Hz level.

	\textbf{c. Change of external magnetic field.}   
	
	To evaluate the impact of the magnetic-field changes on the 1470 nm laser, we measured the frequency drift of the 1470 nm laser by applying an external magnetic field. The laser frequency changed linearly with the magnetic field, with a slope of about 600 kHz/G. Considering the magnetic shielding of the system, we estimated that the linewidth broadening caused by the fluctuations of the external magnetic field was a few Hz, and it would be reduced greatly by designing a better magnetic shielding.  
	
	\textbf{In summary}, due to the influence of the above technical noise, the measured linewidth of the resonant laser and the inhibited laser, which were both 1.2 kHz, were much larger than their quantum-limited linewidths of 150 and 43 Hz, respectively. In addition to the above three types of technical noise, the residual cavity-pulling effect will also cause linewidth broadening. Therefore, the beat-note linewidths when the cavity was resonant and anti-resonant looked the same, as shown in Fig. \ref{Fig3}a.
	
	The temperature fluctuations of the vapor cell were the main factor resulting in the linewidth broadening, which will be further suppressed by cavity-length stabilization and the optimization of the temperature stability. The influence of the pumping-power change and the fluctuations of the residual magnetic field were relatively small. With better power stabilization of the 459 nm laser and the improvement of the temperature precision, we expect that the linewidth of the inhibited laser can be narrowed to the quantum-limit level.

\section*{Data availability}
All the data that support the plots within this paper, including those in the Methods section, are available on Zenodo with the digital object identifier \href{URL}{https://doi.org/10.5281/zenodo.6601106}.

\section*{Code availability}
The codes implementing the calculations of this study are available from the corresponding author upon request.

\bibliographystyle{elsarticle-num}
\bibliography{ref}

	\section*{Acknowledgments}
	
	We acknowledge discussions with C. Peng and Z. Chen. This research was funded by the National Natural Science Foundation of China (NSFC) (91436210), China Postdoctoral Science Foundation (BX2021020), and Wenzhou Major Science \& Technology Innovation Key Project (ZG2020046).

	\section*{Author contributions}
	
	J. C. conceived the idea to use an anti-resonant cavity to realize the inhibited laser as a stable active optical clock. T.S. performed the experiments and carried out the theoretical calculations. T. S. wrote the manuscript. D. P. and J. C. provided revisions. All authors contributed equally to the discussions of the results.

	\section*{Competing interests}
	
	The authors declare no competing interests.

	\section*{Additional information}
	
	%{\bf Extended data} is available for this paper at .
	
	%{\bf Supplementary information} is available for this paper at .
	
	{\bf Correspondence and requests for materials} should be addressed to T. Shi or J. Chen.
	
	%{\bf Reprints and permissions information} is available at .

	%\begin{appendix}
	
	%\end{appendix}

\end{document}